\documentclass{iau} 
\usepackage{graphicx}
\newcommand{\kms}{km~s$^{-1}$}
\newcommand{\gal}{$\alpha$}

\title[LMC Cluster stars] 
{Spectroscopy of LMC Cluster Stars}

\author[Dupree et al.]   
{A. K. Dupree$^1$, C. I. Johnson$^1$, M. Mateo$^2$ and A. P. Milone$^3$}

\affiliation{$^1$Center for Astrophysics $\mid$ Harvard\ \& Smithsonian \\ 60 Garden Street,
Cambridge, MA 02138, USA \\ email: {\tt adupree@cfa.harvard.edu, cjohnson@cfa.harvard.edu} \\[\affilskip]
$^2$Dept. of Astronomy,  University of Michigan, \\ Ann Arbor, Michigan 48109, USA  \\email: {\tt mmateo@umich.edu}\\[\affilskip]
$^3$Dipartimento di Fisica e Astronomia ``Galileo Galilei'', Univ. di Padova, \\Vicola dell'Osservatorio 3,Padova, IT-35122\\email:{\tt antonino.milone@unipd.it}}

\pubyear{2019}
\volume{351}  
\jname{Star Clusters: From the Milky Way to the Early Universe}
\editors{A. Bragaglia, M.B. Davies, A. Sills \& E. Vesperini, eds.}
\begin{document}

\maketitle

\begin{abstract}
High resolution spectra of stars in the $\approx$200 Myr LMC 
globular cluster, NGC 1866, reveal 
rapidly rotating stars with variable H\gal\ emission and absorption, and 
signatures of outflowing material. The variable H\gal\ line can substantially 
affect photometric measurements obtained with HST/WFC3 narrow-band filters.  
 
\end{abstract}

\firstsection 
\section{Introduction}
Photometry from the Hubble Space Telescope has fundamentally changed our
concept of globular clusters (Gratton \etal\ 2012; Bastian and Lardo 2018).  
No longer thought to be 
coeval and chemically  homogeneous, `multiple populations' appear ubiquitous.  We lack
an understanding of the cluster evolution causing this major paradigm shift. 
Many sources for the abundance patterns have been proposed.  However models can not uniquely
identify the source of the chemical enrichments in the Milky Way clusters.  One 
approach is to study  younger clusters in the large Magellanic Cloud (LMC) in the 
hopes of understanding the origins of the multiple populations.

LMC clusters are unusual too. The young and intermediate age
clusters display a unique extended (broadened)  
main sequence turnoff defined by
HST photometry  and a double main sequence (Milone \etal\ 2009, 2018; Goudfrooij \etal\ 2009, 2014). Stellar spectra can reveal physical properties of their members.

\section{Stellar Spectra}
Our spectra were obtained at the {\it Magellan}/Clay 6.5-m telescope at 
Las Campanas Observatory
using two spectrographs: the Magellan Inamori Kyocera Echelle (MIKE) 
and Michigan/{\it Magellan} Fiber System (M2FS, Mateo \etal\ 2012).  MIKE is a high throughput 
double-echelle spectrograph with a blue and red arm spanning the optical 
region: $\sim$~335$-$950 nm (Bernstein \etal\ 2002). 
M2FS is a multi-object spectrograph with 1.2'' fibres and a field of view nearly 30 arcmin in
diameter. For M2FS, we selected a filter spanning 612$-$672 nm, and a slit 
yielding a resolving power
$\lambda/\Delta\lambda \approx$ 28,000.  Targets in the cluster, NGC 1866, were identified
by Milone \etal\ (2017) from the Ultraviolet and Visual Channel of the Wide Field Camera 3
(UVIS/WFC3) of the HST.

\begin{figure}[!ht]
\hspace*{-0.9 cm}
\includegraphics[scale=0.6, angle=90]{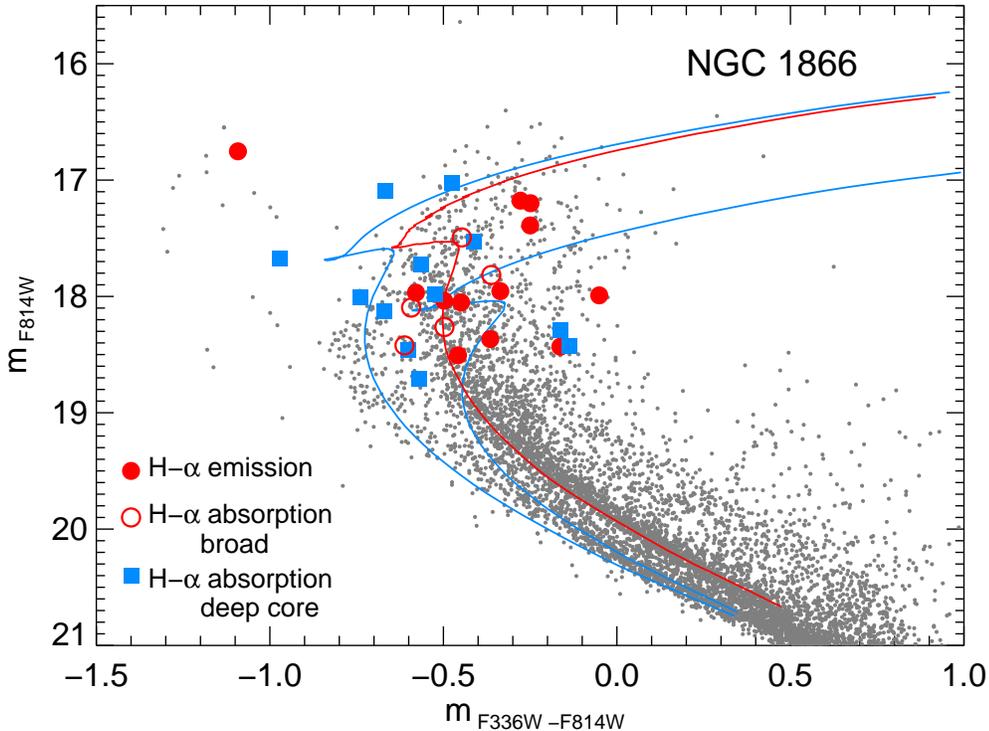} 

 \caption{Color-magnitude diagram (CMD) for NGC 1866 using HST photometry 
(Milone \etal\ 2017). Characteristics
of the H$\alpha$ profiles are marked as derived from high-resolution spectra.
Isochrones for  non-rotating models and ages of 140 Myr and 220 Myr
({\it blue curves}) and one rotating model with an age of 200 Myr ({\it red curve}) are
taken from Georgy \etal\ (2013).  Dots represent cluster stars within 3 arc minutes of the 
cluster center. Figure from Dupree \etal\ (2017), \textcopyright AAS.}   

\end{figure}

\vspace*{-0.3in}
\section{The Color-Magnitude Diagram}
The first spectroscopy (Dupree \etal\ 2017) of turn-off 
stars in a young (200 Myr) LMC cluster, NGC 1866, revealed that rapidly 
rotating stars are indeed present (Figure 1).    The H\gal\ line  appears in 
emission centered on the stellar radial velocity, and is typical of that found in Be stars.
In these  stars, the H\gal\ emission arises in a Keplerian decretion disk
surrounding a rapidly rotating star (Rivinius \etal\ 2013). The emission profile
shape can indicate the orientation of the disk.  Additionally,
other stars exhibit an absorption profile of H\gal\ from which the
stars' rotational velocity ({\it v sin i}) can be inferred.
Stars with emission lie to the
cooler side (the `red')  of the slowly rotating stars in the
HST color magnitude diagram.  Fitting isochrones in Figure 1, suggests that the
population of fast-rotators corresponds to  $\approx$ 200 Myr, and the
blue main sequence may contain two populations of non-rotating stars with
ages of 140 and 220 Myr.  Marino \etal\ (2018) has measured the
projected rotational velocities of main sequence stars from the H\gal\ and He~I profiles 
in a very young ($\approx$40 Myr) LMC  
cluster, NGC 1818.  They 
find the blue main sequence to have a lower mean rotation (71$\pm$10 \kms)
than the red main sequence (202$\pm$23 \kms), confirming that rotation
is a significant parameter in defining the appearance of the CMD. 

\vspace*{-0.2in}
\section{Spectroscopic Variability}
It is well known that the H\gal\ emission can vary in Be stars,
as studies of Milky Way Be stars have shown (Dimitrov \etal\ 2018). Not surprisingly, the LMC
cluster stars vary in their H\gal\ profiles too (Fig. 2, 3). In addition 
to radial velocity shifts, this can be of
particular concern for HST/WFC3 photometry which  employs a narrow-band
filter, F656N, with a passband of only $\approx$18\AA\ (6552.6-6570.6\AA, full
width at half power) to detect H\gal\ emission.  Radial velocities of H I in the LMC
can vary between +200 to +330 \kms\ (Staveley-Smith \etal\ 2003)
such that the H\gal\ profile will not be captured (or completely covered)
by the filter.  Moreover the variability 
can amount to a factor of two or more in the flux which  creates
scatter in a CMD using this photometric parameter. 

\begin{figure}
\hspace*{-0.9 cm}
\includegraphics[scale=0.6]{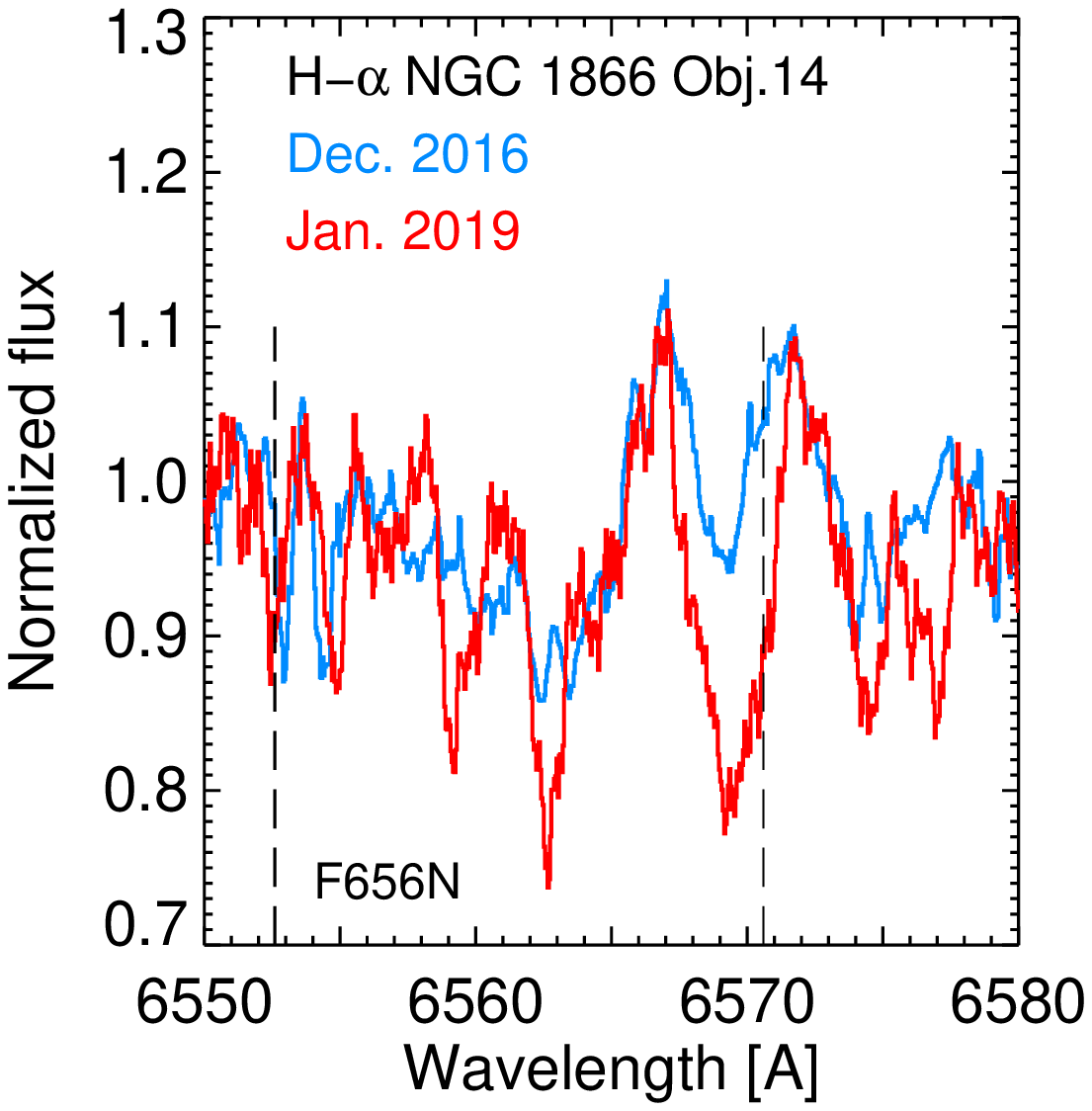}

\vspace*{-3in}

\hspace*{+2.2in}
\includegraphics[scale=0.6]{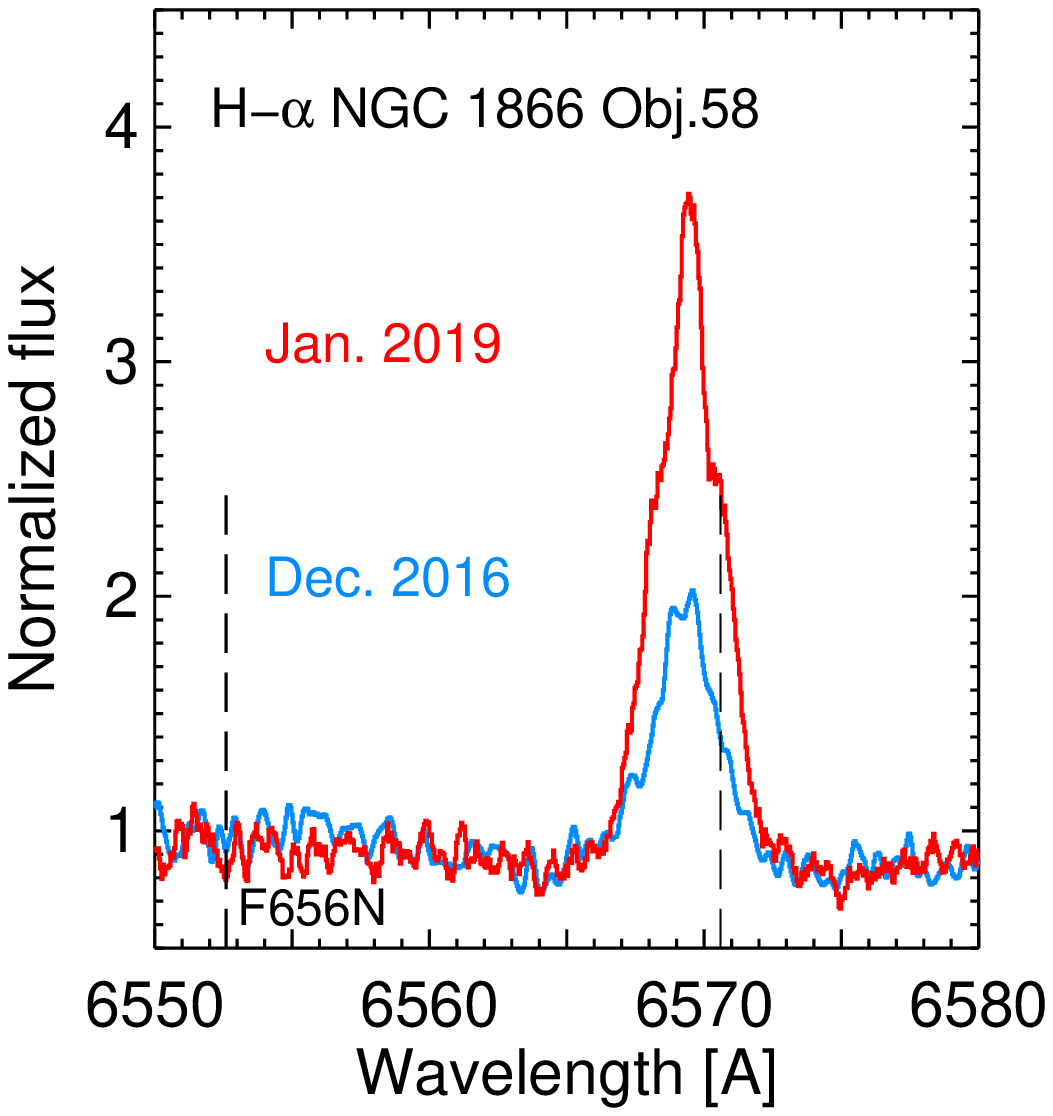}

\caption{Two stars exhibiting substantial H\gal\ variability. (Identifications in 
Dupree \etal\ 2017).  The broken
lines mark the full width at half power  of the response of the  
F656N filter on HST/WFC3. Note  that the LMC cluster has a substantial
velocity shift (v$_{helio}$= $+$298.5 \kms, Mucciarelli \etal\ 2011) such that the 
filter does  not capture the full H\gal\ line profile.} 

\end{figure}

\vspace*{-0.2in}
\section{Braking and Spin-Down}

The presence of a split main sequence consisting of rapidly
rotating and slowly rotating stars in LMC clusters,  led to the suggestion 
(D'Antona \etal\ 2017) that
cluster stars may be formed initially in a rapidly rotating state. 
Subsequently they are braked by a stellar wind or as a result
of low-frequency oscillation modes if  in a binary system, and spun down to become the
slowly-rotating stars on the blue main sequence. Such a 
phenomenon would minimize, if not remove,  the apparent age difference between the
fast and slowly-rotating stars and perhaps explain the extended 
main-sequence turnoff in these clusters. Although a scenario invoking a combination
of rotation and a distribution of ages may still be required (Goudfrooij \etal\ 2017). 
In fact, rotational braking has been inferred in a B-dwarf star, 
$\sigma$ Ori E, based on photometric measures of a systematic change in the
star's rotation rate (Townsend \etal\ 2010). If mass outflow is the culprit causing
the braking, one could seek spectroscopic
signatures of such outflow directly from the line profiles. Ultraviolet
profiles clearly indicate accelerating outflows occur in Be stars with 
terminal velocities ranging from $\sim -$500 to $-$1200 \kms\ (Slettebak 1994). 
It may be possible to detect the onset of such flows in Balmer
series profiles marked by excess absorption on the `blue' wing caused
by outflowing gas as appears in the pole-on Obj 58 (Figure 3).  This may
be a special object as rapidly rotating stars are thought to
have a ratio of polar to equatorial mass loss rate of a 
factor of 40 (Krti\u{c}ka 2014).  Further observations at high signal-to-noise
would be useful.

\begin{figure}
\hspace*{-0.8 cm}
\includegraphics[scale=0.6]{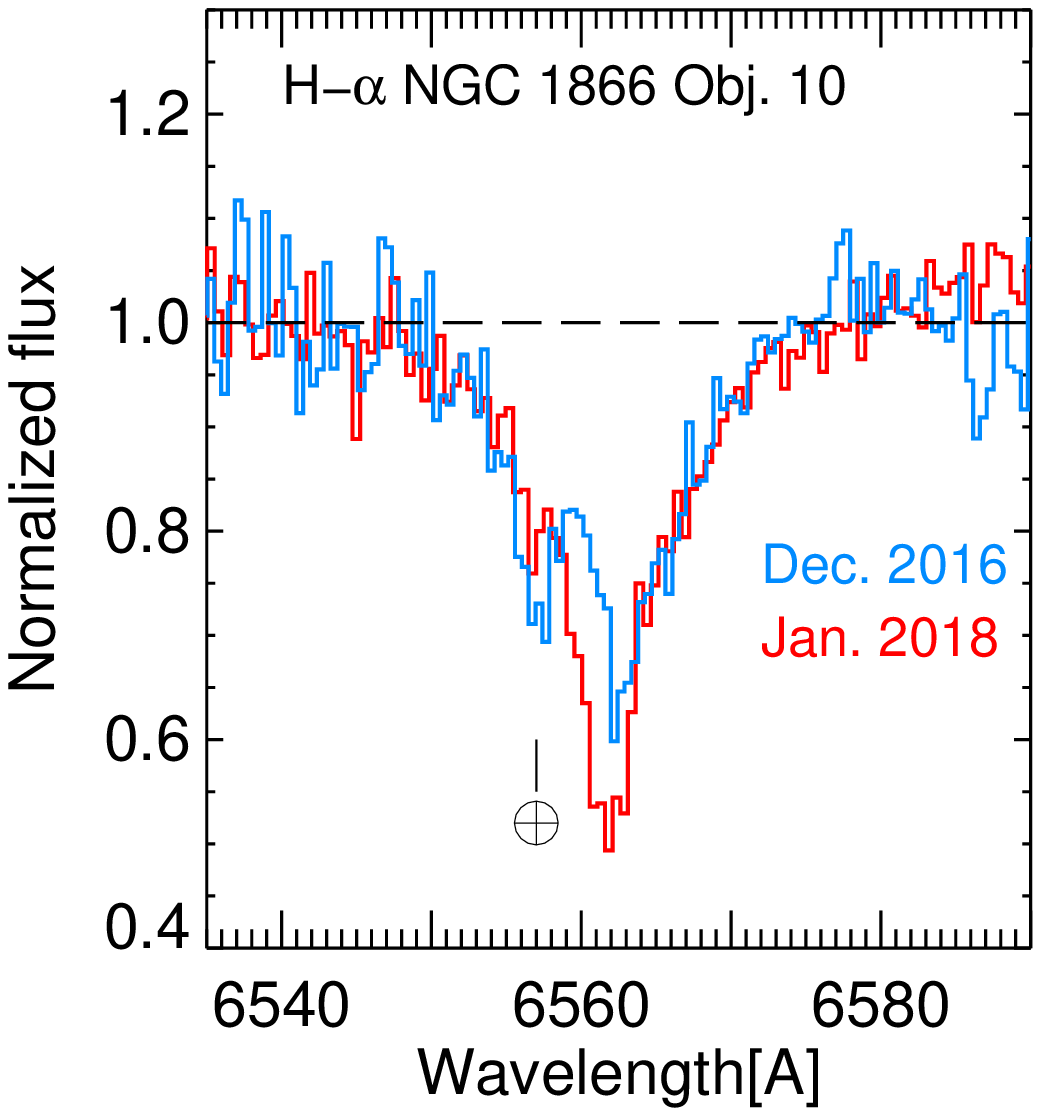}

\vspace*{-3. in}

\hspace*{+2.3 in}
\includegraphics[scale=0.6]{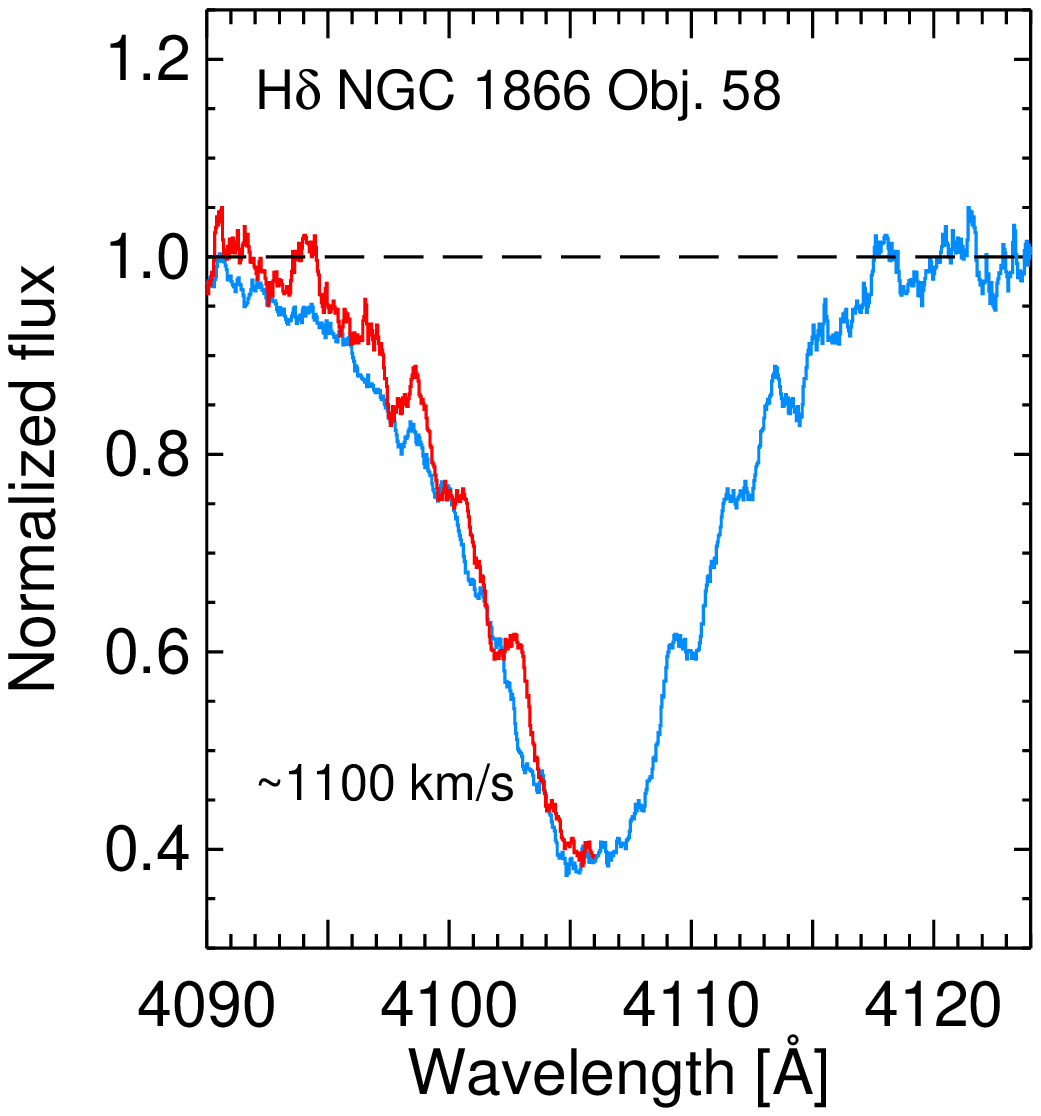}

\caption{\textit{Left:} The variable H\gal\ line.  \textit{Right:} The H$\delta$ 
transition in a rapidly rotating pole-on star. The long wavelength wing has
been centrally reversed ({\it red curve}) over the short wavelength wing demonstrating
increased opacity on the short wavelength side,  likely due to an outflow extending to $\approx$1100 \kms.}    

\end{figure}

\end{document}